# Ferromagnetism as a universal feature of nanoparticles of the otherwise nonmagnetic oxides

A. Sundaresan*, R. Bhargavi, N. Rangarajan, U. Siddesh and C. N. R. Rao

*Chemistry and Physics of Materials Unit and Department of Science and Technology Unit on Nanoscience, Jawaharlal Nehru Centre for Advanced Scientific Research, Jakkur P. O., Bangalore 560 064 India.*

## Abstract

Room-temperature ferromagnetism has been observed in the nanoparticles (7 - 30 nm dia) of nonmagnetic oxides such as $CeO_2$, $Al_2O_3$, ZnO, $In_2O_3$ and $SnO_2$. The saturated magnetic moments in $CeO_2$ and $Al_2O_3$ nanoparticles are comparable to those observed in transition metal doped wide band semiconducting oxides. The other oxide nanoparticles show somewhat lower values of magnetization but with a clear hysteretic behavior. Conversely, the bulk samples obtained by sintering the nanoparticles at high temperatures in air or oxygen became diamagnetic. As there were no magnetic impurities present, we assume that the origin of ferromagnetism may be due to the exchange interactions between localized electron spin moments resulting from oxygen vacancies at the surfaces of nanoparticles. We suggest that ferromagnetism may be a universal characteristic of nanopartilces of metal oxides.

Integration of semiconductor with ferromagnetic functionality of electrons has been the focus of recent research in the area of spintronics because of the difficulties associated with the injection of spins into nonmagnetic semiconductors in conventional spintronic devices. Ferromagnetism in semiconductors and insulators are rare, the well known ferromagnetic semiconductors being the chalcogenides, EuX (X = O, S and Se) ($T_C$< 70 K) and $CdCr_2X_4$ (X = S and Se) ($T_C$< 142 K ) with the rock salt and spinel structure respectively.[1,2] Following the theoretical prediction of Dietl et al. that Mn doped ZnO and GaN could exhibit ferromagnetism above room temperature,[3] several studies have focused on films and bulk samples of metal oxides such as $TiO_2$, ZnO, $In_2O_3$, $SnO_2$ and $CeO_2$ doped with Mn, Co and other transition metal ions.[4-8]

While the existence of ferromagnetism in transition metal doped semiconducting oxides remains controversial,[9] thin films of the band insulator $HfO_2$ have been reported to exhibit ferromagnetism at room temperature in the absence of any doping.[10] This is puzzling, since pure $HfO_2$ does not have any magnetic moment and the bulk sample is diamagnetic. Similar ferromagnetism has been reported in other nonmagnetic materials such as $CaB_6$, CaO and SiC where the origin of ferromagnetism is believed to be due to intrinsic defects.[11-13] It has been suggested that ferromagnetism in thin films of $HfO_2$ may be related to anion vacancies.[14] It has been reported very recently that thin films of undoped $TiO_2$ and $In_2O_3$ also show ferromagnetism at room temperature, [15] the corresponding bulk forms of these materials being diamagnetic. Thin films of these oxides might have defects or oxygen vacancies which could be responsible for the observed ferromagnetism. *Ab initio* electronic structure calculations using density functional theory in $HfO_2$ have shown that isolated halfnium vacancies lead to ferromagnetism.[16] Meanwhile, there is a conflicting report attributing the ferromagnetism in $HfO_2$ to possible iron contamination while using stainless-steel tweezers in handling thin films.[17]

In this Rapid communication, we report the discovery of ferromagnetism at room temperature in nanoparticles of nonmagnetic oxides such as $CeO_2$, $Al_2O_3$, ZnO, $In_2O_3$ and $SnO_2$. Our studies show that ferromagnetism is associated only with the nanoparticles while the corresponding bulk samples are diamagnetic. The origin of ferromagnetism in these materials is assumed to be the exchange interactions between localized electron spin moments resulting from the oxygen vacancies at the surfaces of the nanoparticles.

Nanoparticles of $CeO_2$, $Al_2O_3$, ZnO, $In_2O_3$ and $SnO_2$ were prepared by the methods described in the literature.[18-20] The preparation methods do not involve any magnetic element and therefore we rule out the possibility of contamination of magnetic impurities. For example, the nanoparticles of $CeO_2$ were prepared by the addition of hexamethylenetetramine to a solution of cerium nitrate $[Ce(NO_3)_3]$ under constant stirring.[18] The nanoparticles of all these oxides were annealed at temperatures between 400 °C and 500 °C in flowing oxygen to remove organic matters. In order to make bulk samples, these nanoparticles were sintered at high temperatures (1000 – 1400 °C). Powder x-ray diffraction (XRD) was used to identify the phase, its purity and to determine the grain size. The particle size and morphology were studied by Field Emission Scanning Electron Microscopy (FESEM) and Transmission Electron Microscopy (TEM). Magnetization measurements were carried out with vibrating sample magnetometer in physical property measuring system (PPMS, Quantum Design, USA).

XRD patterns of all the samples showed that they were monophasic with broad peaks characteristic of the nanoparticles. The lattice parameters and the full-width at half-maximum of all the reflections were obtained from the Rietveld refinement in the pattern matching mode using the program **FULLPROF**.[21] The lattice parameters of the oxide nanoparticles were generally higher than those of the corresponding bulk forms. For example, the lattice parameter of the $CeO_2$ nanoparticles (7 nm) is 5.424(3) Å whereas that of the corresponding bulk sample is 5.413(1) Å. This is in agreement with the earlier report that the lattice expands in oxide nanoparticles.[22] The increase of lattice with decreasing particles size might results from the oxygen vacancy associated with nanoparticles. Similar results were obtained for $Al_2O_3$, ZnO, $In_2O_3$ and $SnO_2$ samples. The average particle sizes of $CeO_2$, $Al_2O_3$, ZnO, $In_2O_3$ and $SnO_2$ estimated by the Scherrer's formula using all diffraction lines were 15, 4, 30, 12 and 20 nm respectively.

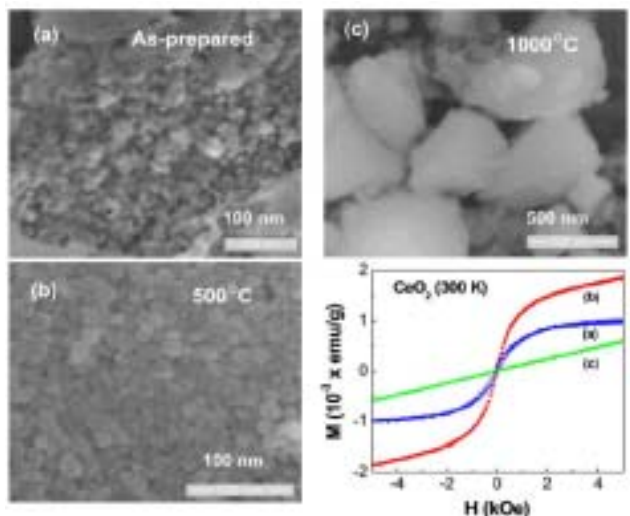

FIG.1. FESEM images of $CeO_2$ nanoparticles [(a) 7 nm, (b) 15 nm, (c) 500 nm] and their magnetization curves at 300 K. Note the absence of ferromagnetism in the 500 nm nanoparticles, in contrast to the 7 or 15 nm nanoparticles.

In Fig. 1 we show the room-temperature magnetization-field curves of many $CeO_2$ samples: (a) as-prepared (b) heated at 500 °C for one hour and (c) heated at 1000 °C for one hour. We have shown the FESEM images of these three samples in the figure. It can be seen that the as-prepared particles (7 nm) are covered by the organic coating used in the preparation of nanoparticles whereas the 500 °C heated particles (15 nm) are free from such coating. It is obvious from the *M(H)* curves that the as-prepared and 500 °C heated nanoparticles show ferromagnetic behavior with coercivity ~100 Oe. This is surprising, since the bulk $CeO_2$ is a band insulator with $Ce^{4+}$ in the $4f^0$ electronic configuration. On the other hand, the ferromagnetism is suppressed in the 1000 °C sample with ~500 nm size particles and this sample exhibits a linear M(H) behavior with low magnetic moment, a behavior close to diamagnetism as normally expected of $CeO_2$.

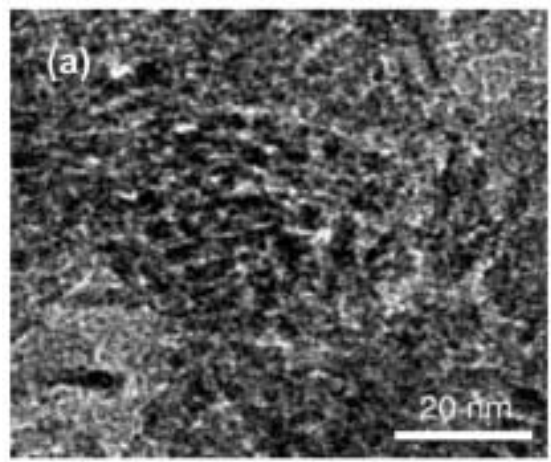

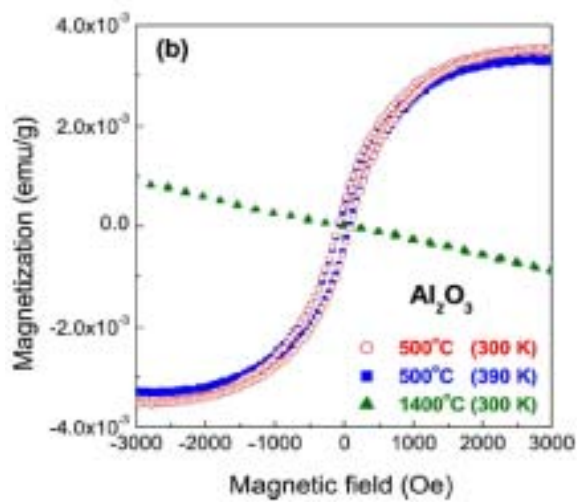

FIG. 2. (a) TEM image of $Al_2O_3$ nanoparticles heated at 500 °C and (b) their magnetization curves showing ferromagnetism even at 390 K. Note that the nanoparticles sintered at 1400 °C exhibit diamagnetic behavior at 300 K.

A TEM image of $Al_2O_3$ nanopartilces obtained by heating Al (OH)$_3$ at 500 °C is shown in Fig. 2(a). The *M(H)* curves of these nanoparticles (0.0291 g) recorded at 300 K and 390 K are shown in Fig. 2(b). These nanoparticles show ferromagnetism even at 390 K with clear hysteretic behavior. The saturation magnetic moment at 300 K is ~3.5 X $10^{-3}$emu/g, comparable to that reported for Mn-doped ZnO.[5] In order to verify that the room-temperature ferromagnetism is associated only with nanoparticles, the nanoparticles of the sample were pressed into a bar and sintered at 1400 °C for one hour in air to obtain bulk samples with micron-sized particles. The magnetization of the bulk sample thus obtained is shown in Fig. 2(b). It is clear form this figure that the bulk sample is diamagnetic. Similarly, room-temperature ferromagnetism is observed in ZnO nanoparticles heated at 400 °C and diamagnetic behavior in the sample sintered at 1200 °C (Fig. 3).

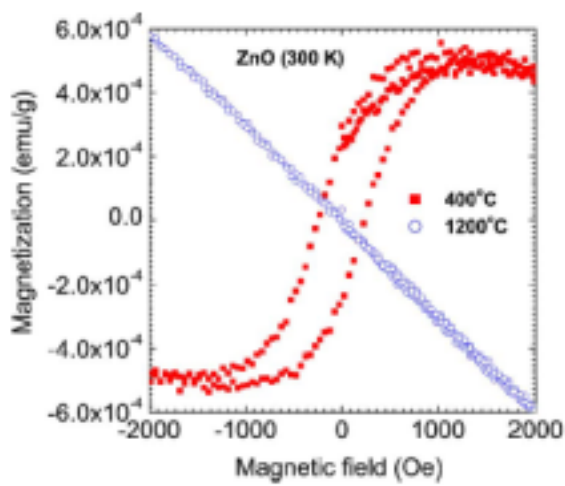

FIG. 3. *M* versus *H* curves measured at 300 K for nanoparticles of ZnO heated at 400 °C and sintered at 1200 °C.

Unlike $CeO_2$, $Al_2O_3$ and ZnO which are insulators, $In_2O_3$ and $SnO_2$ are transparent conductors with a wide and gap (~3.6 eV). Magnetization data of $In_2O_3$ and $SnO_2$ nanoparticles are shown in Fig. 4. The magnetization behavior of $SnO_2$ is slightly different from that of the other oxides, but similar to that observed in thin films of Co doped $SnO_2$.[7] It can be seen from this figure that there is a small hysteresis at low fields and that the magnetic moment increases linearly at higher field. The linear behavior may be due to magnetic moments associated with conduction electrons. This is consistent with the observation that the nanoparticles after sintering at 1200 °C show paramagnetic behavior. Though there may be slight differences in the magnetization behavior, nanoparticles of all the oxides studied exhibit room-temperature ferromagnetism. It should be noticed that the nanoparticles of paramagnetic metallic $ReO_3$ with low magnetic susceptibility are reported to show hysteresis at 5 K.[24] As the magnetic susceptibility of $ReO_3$ nanoparticles is relatively low, it may show ferromagnetism even at room temperature.

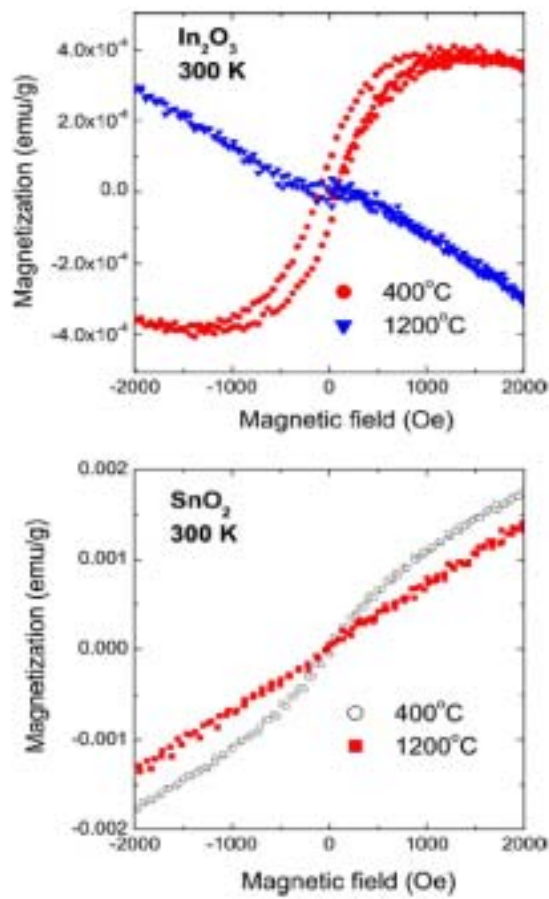

FIG. 4. *M* versus *H* curves measured at 300 K for nanoparticles of $In_2O_3$ and $SnO_2$ heated at 400 °C and sintered at 1200 °C.

The origin of ferromagnetism in the nanoparticles of these nonmagnetic oxides seems to be similar to that in the thin films of $HfO_2$, $TiO_2$ and $In_2O_3$ where the oxygen deficiency results from thin film growth conditions.[10, 15] In contrast to thin films, where the contamination of films by handling can vitiate the results, ferromagnetism in the oxide nanoparticles is robust and universal. We suggest that the unpaired electron spins responsible for ferromagnetism in the nanoparticles have their origin in the oxygen vacancies, especially on the surfaces of the oxide nanoparticles. The nature of exchange interactions between them is not clear at present. However, onemay expect that electrons trapped in oxygen vacancies ($F$ center) are polarized to give room-temperature ferromagnetism. This mechanism has been proposed to explain ferromagnetism in some transparent oxides.[23]

In conclusion, we have shown that nanoparticles of metal oxides such as $CeO_2$, $Al_2O_3$, ZnO, $In_2O_3$ and $SnO_2$, exhibit room-temperature ferromagnetism whereas the corresponding bulk oxides exhibit diamagnetism. We assume that the origin of ferromagnetism may be due to the exchange interactions between unpaired electron spins arising from oxygen vacancies at the surfaces of the nanopartilces. We suggest that all metal oxides in nanoparticulate form would exhibit room-temperature ferromagnetism. The ferromagnetism assumed to be associated with oxygen vacancies gives a possible clue to understand some of the contradicting findings in the dilute magnetic semiconducting oxides.

The authors thank R. V. K. Mangalam, Chandra Sekhar Rout, and C. Madhu for their help in the sample preparation and magnetic measurements. The authors Bhargavi, Rangarajan and Siddesh would like to thank JNCASR for providing opportunity to do research work under the program Project Oriented Chemical Education (POCE) and Summer Research, respectively. This work was supported by Department of Science and Technology, India under the nanoscience initiative program.

*Electronic address: sundaresan@jncasr.ac.in